\documentclass[twocolumn,aps,prb]{revtex4}
\usepackage{array}
\usepackage{amssymb}
\usepackage{amsmath}
\usepackage{graphicx}
\usepackage{multirow}
\usepackage{color}
\usepackage{comment}
\usepackage{url}

\begin{document}

\title{Colored-noise thermostats {\em {\`a} la carte}\footnote{Reprinted with permission from J. Chem. Theory Comput., 2010, 6 (4), pp 1170–1180. Copyright 2010 American Chemical Society.}}
\author{Michele Ceriotti}
\email{michele.ceriotti@phys.chem.ethz.ch}
\affiliation{Computational Science, Department of Chemistry and Applied Biosciences,
ETH Z\"urich, USI Campus, Via Giuseppe Buffi 13, CH-6900 Lugano, Switzerland}
\author{Giovanni Bussi}
\affiliation{Universit\`a di Modena e Reggio Emilia and INFM-CNR-S3, via Campi 213/A, 41100 Modena, Italy}
\author{Michele Parrinello}
\affiliation{Computational Science, Department of Chemistry and Applied Biosciences,
ETH Z\"urich, USI Campus, Via Giuseppe Buffi 13, CH-6900 Lugano, Switzerland}
\date{\today}

\begin{abstract}
Recently, we have shown how a colored-noise Langevin equation can be used
in the context of molecular dynamics as a tool to obtain dynamical trajectories whose 
properties are tailored to display desired sampling features.
In the present paper, after having reviewed some analytical results 
for the stochastic differential equations forming the basis of our approach, we describe in detail
the implementation of the generalized Langevin equation thermostat and the fitting procedure used to
obtain optimal parameters. We discuss in detail the simulation of nuclear quantum effects, 
and demonstrate that, by carefully choosing parameters, one can successfully model strongly anharmonic 
solids such as neon. For the reader's convenience, a library of thermostat parameters
and some demonstrative code can be downloaded from an on-line repository.
\end{abstract}
\maketitle

Stochastic differential equations (SDE) have been used to model the time evolution 
of processes characterized by random behavior, in  fields as diverse as
physics and economics. In particular, the Langevin equation (LE)
has been regularly applied in the study of Brownian motion,
and has been used extensively in molecular dynamics (MD) computer simulations
as a convenient and efficient tool to obtain trajectories which sample the constant-temperature,
canonical ensemble\cite{schn-stol78prb,adel-broo82jpc}.

In its original form, the Langevin equation is based on the 
assumption of instantaneous system-bath interactions, which correspond to the 
values of the random force being uncorrelated at different times. 
A non-Markovian, generalized version of the LE arises
in the context of Mori-Zwanzig theory\cite{zwan61pr,zwan+01book}.
In this theory, if one considers a harmonic system coupled with a harmonic bath,
it is possible to integrate out the degrees of freedom
of the bath. This leaves one with a linear stochastic equation where both the 
friction and the noise have a finite memory. The conventional LE is 
recovered in the limit of a clear separation between the characteristic 
time-scale of the system's dynamics and that of the system-bath interaction.

This class of non-Markovian SDEs has been extensively used to model the 
dynamics of open systems interacting with a physically-relevant bath
(see e.g.~Refs.~\cite{mart02jcp,wang07prl,kant08prb}).
Instead, our recent works\cite{ceri+09prl,ceri+09prl2} have used colored(correlated)-noise SDEs
as a device to sample efficiently statistical distributions
in molecular-dynamics (MD) simulations.
These works aimed to show how a
stochastic thermostat suitable for Car-Parrinello-like dynamics\cite{ceri+09prl}
could be constructed, and to include nuclear quantum effects in a large class of problems
at a fraction of the cost of path-integrals calculations\cite{ceri+09prl2}. 
In these applications the real dynamics is lost, and one focuses only on the efficient calculation
of static ensemble averages.

In this Paper we discuss the practical implementation of the generalized Langevin equation (GLE)
thermostat that we used in the two cases mentioned above.
We also provide the reader with the analytical and numerical tools needed
to construct SDEs tailored to their own sampling needs.
Throughout we take advantage of the dimensional reduction
scheme, which allows to exploit the equivalence between a non-Markovian
dynamics and a Markovian dynamics in higher dimensionality.
In doing this, we supplement the physical coordinates with additional degrees of 
freedom\cite{zwan+01book}, whose equations of motion are taken as linear, so as 
to simplify the formalism  and analytical derivations.

In the Appendices we recall some of the properties of multidimensional
stochastic processes\cite{gard03book,lucz05chaos,mart02jcp,marc-grig83jcp},
 which are useful to our discussion, and present a short
comparison of the GLE thermostat and the widely used massive Nos\'e-Hoover chains\cite{nose84jcp,hoov85pra,mart+92jcp,tobi+93jpc}.
A simple FORTRAN90 code implementing our method to the dynamics of an harmonic oscillator and a library
of optimized thermostat parameters can be downloaded from an on-line repository\cite{PARS}.


\section{Generalized Langevin thermostat\label{sec:gle}}
\subsection{Markovian and non-Markovian formulations}
The Langevin equation for a particle with position $q$ and momentum $p$, subject to a 
potential $V(q)$, can be written as
\begin{equation}
 \begin{split}
\dot q=&p\\
\dot p=&-V'(q) -a_{pp} p + b_{pp}\xi(t). 
 \end{split}\label{eq:langevin}  
\end{equation}
where $\xi(t)$ represent an uncorrelated, Gaussian-distributed random force 
with unitary variance and zero mean [$\left<\xi\right>=0$, $\left<\xi(t)\xi(0)\right>=\delta(t)$]. 
Here and what follows we use mass-scaled coordinates. Furthermore, for consistency, the friction 
coefficient (usually denoted by $\gamma$) is here given the symbol $a_{pp}$, 
while $b_{pp}$ is the intensity of the random force. 
In this notation, the fluctuation-dissipation theorem (FDT) reads
$b_{pp}^2= 2 a_{pp} k_B T$. If this relation holds, the dynamics generated 
by Eq.~(\ref{eq:langevin}) will sample the canonical ensemble at temperature $T$\cite{kubo66rpp,zwan+01book}.

As explained in the Introduction, in order to bypass the complexity of dealing with
a non-Markovian formulation directly, we supplement the system with
$n$ additional degrees of freedom  $\mathbf{s}=\left\{s_i\right\}$
which are linearly coupled  to the 
physical momentum and between themselves.
The resulting SDE can be cast into the compact form
\begin{equation}
\begin{split}
  \dot{q}=&p\\
\!\left(\! \begin{array}{c}\dot{p}\\ \dot{\mathbf{s}} \end{array}\!\right)\!=&
\left(\!\begin{array}{c}-V'(q)\\ \mathbf{0}\end{array}\!\!\right)
\!-\!\left(\!
\begin{array}{cc}
a_{pp} & \mathbf{a}_p^T \\ 
\bar{\mathbf{a}}_p & \mathbf{A}
\end{array}\!\right)\!
\left(\!\begin{array}{c}p\\ \mathbf{s}\end{array}\!\right)\!+\!
\left(\!
\begin{array}{cc}
b_{pp} & \mathbf{b}_p^T \\ 
\bar{\mathbf{b}}_p & \mathbf{B}
\end{array}\!\right)\!
\left(\!\begin{array}{c}\multirow{2}{*}{$\boldsymbol{\xi}$}\\ \\\end{array}\!\right),
\end{split}
\label{eq:mark-sde}
\end{equation}
Here, $\boldsymbol{\xi}$ is a vector of $n+1$ uncorrelated Gaussian random numbers, with
$\left<\xi_i\left(t\right)\xi_j\left(0\right)\right>=\delta_{ij}\delta\left(t\right)$.
Clearly, Eq.~(\ref{eq:langevin}) is recovered when $n=0$. 
For an harmonic potential $V(q)=\frac{1}{2}\omega^2 q^2$, Eqs.~(\ref{eq:mark-sde}) are linear,
and an Ornstein-Uhlenbeck process is recovered whose time propagation
can be evaluated analytically.
In the non-linear case one can use the Trotter-decomposition to
split the dynamics into a linear part, which evolves the $(p,\mathbf{s})$ momenta,
and a non-linear part, which evolves the Hamilton equations\cite{buss-parr07pre}.
This is facilitated by the fact that the dynamics of $(p,\mathbf{s})$ alone
is linear, and its exact finite-time propagator can be analytically
evaluated (see Subsection~\ref{sub:implementation}).

Here and in the rest of the paper, we adopt the same notation introduced in Ref.~\cite{ceri+09prl2} 
to distinguish between matrices acting on the full state vector 
$\mathbf{x}=\left(q,p,\mathbf{s}\right)^T$ or on parts of it, as illustrated below:
\newcommand\arS{\rule{0pt}{12pt}}
\begin{equation}
\begin{array}{ccccc}
      &   q   &    p   &   \mathbf{s}  & \arS \\ \cline{2-4}
\multicolumn{1}{c|}{q} & m_{qq} & m_{qp} & \multicolumn{1}{c|}{\mathbf{m}_q^T} & \arS \\\cline{3-4}
\multicolumn{1}{c|}{p} & \multicolumn{1}{c|}{\bar{m}_{qp}} &  m_{pp} &  \multicolumn{1}{c|}{\mathbf{m}_p^T} & \arS \\\cline{4-4}
\multicolumn{1}{c|}{\mathbf{s}} &  \multicolumn{1}{c|}{\bar{\mathbf{m}}_q}  &  \multicolumn{1}{c|}{\bar{\mathbf{m}}_p} &  \multicolumn{1}{c|}{\mathbf{M}} & \arS \\\cline{2-4}
\end{array}
\hspace{-8pt}\begin{array}{cc}
\arS \\ \arS \\
\left.\rule{0pt}{12pt}\right\}\!\mathbf{M}_p \\
\end{array}
\hspace{-8pt}\begin{array}{cc}
\arS \\
\left.\rule{0pt}{20pt}\right\}\!\mathbf{M}_{qp} 
\end{array}
\label{eq:notation}
\end{equation}

The Markovian dynamical equations~(\ref{eq:mark-sde}) are equivalent  to
a non-Markovian process for the physical variables only.
This is best seen by first considering the evolution of the 
$(p,\mathbf{s})$ variables in the free-particle analogue of Eqs.~(\ref{eq:mark-sde}).
The additional degrees of freedom $\mathbf{s}$ can be integrated away, and one is left
with (see Ref.~\cite{zwan+01book} and Appendix~\ref{app:memory})  
\begin{equation}
 \dot{p}=-\int_{-\infty}^t K(t-s) p(s)\mathrm{d} s +\zeta(t)
\end{equation}
where the memory kernel $K(t)$ is related to the elements of $\mathbf{A}_p$ by
\begin{equation}
K(t)=2a_{pp} \delta(t)-\mathbf{a}_p^T e^{-\left|t\right|\mathbf{A}}\bar{\mathbf{a}}_p. \label{eq:kernel-k}
\end{equation}
Based on the fact that the the free-particle dynamics of $(p,\mathbf{s})$ is an
OU process, one also finds than the relationship between the static covariance matrix
$\mathbf{C}_p=\left< \left(p,\mathbf{s}\right)^T \left(p,\mathbf{s}\right)\right>$,
the drift matrix $\mathbf{A}_p$ and the diffusion matrix $\mathbf{B}_p$ is given by:\footnote{Note the remarkable
formal analogy between Eq.~(\ref{eq:free-cov}) and the equations for the orthogonality constraints
in Car-Parrinello dynamics, see e.g. Ref.~\cite{marx-hutt00proc}}
\begin{equation}
\mathbf{A}_p\mathbf{C}_p+\mathbf{C}_p\mathbf{A}_p^T=\mathbf{B}_p\mathbf{B}_p^T.
\label{eq:free-cov}
\end{equation}
In Appendix~\ref{app:memory} we show that setting $\mathbf{C}_p=k_B T$ 
is sufficient to satisfy the FDT. In this case, 
Eq.~(\ref{eq:free-cov}) fixes $\mathbf{B}_p$ once
$\mathbf{A}_p$ is given. FDT also implies that the 
colored-noise autocorrelation function $H(t)=\left<\zeta(t)\zeta(0)\right>$ 
is equal to $k_BTK(t)$, whereas the more complex relation between $K(t)$
and $H(t)$, valid in the general case, is reported in Eq.~(\ref{eq:dred-mem-t}).

Since there is no explicit coupling between the position $q$ and the
additional momenta $\mathbf{s}$, one can check that exactly the same dimensional
reduction can be performed in the case of an arbitrary potential coupling
$p$ and $q$, and that Eqs.~(\ref{eq:mark-sde}) correspond to the non-Markovian process
\begin{equation}
\begin{split}
 \dot{q}&=p\\
 \dot{p}&=-\frac{\partial V}{\partial q}-\int_{-\infty}^t K(t-s) p(s)\mathrm{d} s +\zeta(t).
\end{split}
\label{eq:nonmark-sde}
\end{equation}
In the memory kernel~(\ref{eq:kernel-k}), $\mathbf{A}$ can be chosen to 
be a general real matrix, and can have complex eigenvalues,
provided they have a positive real part. 
This results in a $K(t)$ that is a linear combination of exponentially damped 
oscillations. Therefore, a vast class of non-Markovian dynamics can be represented
by Markovian equations such as~(\ref{eq:mark-sde}).

\subsection{Exact solution in the harmonic limit}
The thermostats typically used in MD simulations have  a few parameters, that are 
chosen by trial and error.
A thermostat based on Eqs.~(\ref{eq:mark-sde}) depends on a much larger number of 
parameters, and hence the fitting procedure is more complex.  It is therefore important to 
find ways to compute {\em a priori} analytical estimates so as  to guide the tuning
of the thermostat.

To this end, we examine the harmonic oscillator, which is commonly used to 
model physical and chemical systems.
By choosing $V(q)=\frac{1}{2}\omega^2 q^2$ the force term in~(\ref{eq:mark-sde}) 
becomes linear, and the dynamics of $\mathbf{x}=(q,p,\mathbf{s})^T$ 
is the OU process $\dot{\mathbf{x}}=-\mathbf{A}_{qp}\mathbf{x}+\mathbf{B}_{qp}\boldsymbol{\xi}$.
In Eqs.~(\ref{eq:mark-sde}) the $\mathbf{s}$ degrees of freedom are coupled to the 
momentum only. Therefore, most of the additional entries in $\mathbf{A}_{qp}$ and $\mathbf{B}_{qp}$
are zero, and the equations for $\mathbf{x}$ read
\begin{equation}
\!\left(\!\begin{array}{c}\dot{q} \\\dot{p}\\ \dot{\mathbf{s}}\end{array}\!\right)\!=
-\!\left(\!
\begin{array}{ccc}
0 & -1 & \mathbf	{0} \\
\omega^2 & a_{pp} & \mathbf{a}_p^T \\ 
\mathbf{0} & \bar{\mathbf{a}}_p & \mathbf{A}
\end{array}\right)
\!\left(\!\begin{array}{c}q\\ p\\ \mathbf{s}\end{array}\!\right)\!+
\!\left(\!\begin{array}{ccc}
0 & 0& \mathbf{0}\\
0 & \multicolumn{2}{c}{\multirow{2}{*}{$\mathbf{B}_p$}}\\
\mathbf{0} & & \\
\end{array}\!\right)\!
\!\left(\!\begin{array}{c}0\\\multirow{2}{*}{$\boldsymbol{\xi}$} \\ \\\end{array}\!\right)\!
\label{eq:mark-pq}.
\end{equation}

The exact finite-time propagator for Eqs.~(\ref{eq:mark-pq}) can be computed, and so it is possible 
to obtain any ensemble average or time-correlation function analytically. 
Of course, one is most interested in the expectation values of the physical variables $q$ and $p$. 
In particular, one can obtain the fluctuations $\left<q^2\right>$ and $\left<p^2\right>$
and correlation functions of the form $\left<q^2(t)q^2(0)\right>$, which can be used to measure the 
coupling between the thermostat and the system. The resulting expressions are simple to evaluate but lengthy,
and we refer the reader to Appendix~\ref{app:corr-time} for their explicit form.

One can envisage, using the estimates computed for an oscillator of frequency $\omega$,
to predict and hence optimize the response of a normal mode of a similar frequency in the 
system being studied. Furthermore, thanks to the properties of Eq.~(\ref{eq:mark-pq}), one does not need to perform 
a normal-modes analysis to turn this idea into a practical method. 
Consider indeed a perfect harmonic crystal, and apply an independent instance of the 
GLE thermostat to the three cartesian coordinates of each atom. It is easy to see that, 
since Eq.~(\ref{eq:mark-pq}) is linear, and contains Gaussian noise, the thermostatted 
equations of motion are invariant under any orthogonal transformation of the coordinates. 
Therefore, the resulting dynamics can be described on the basis of the normal
modes just as in ordinary Hamiltonian lattice dynamics. As a consequence, each phonon 
will respond independently as a 1-D oscillator with its own characteristic frequency.
Thus, to tune the GLE thermostat, one only needs the analytical results in the one-dimensional 
case, evaluated as a function of $\omega$. The parameters can then be optimized for a number of different purposes,
based solely on minimal information on the vibrational spectrum of the system under investigation, 
without any knowledge of the phonons eigenmodes.

The invariance properties of the GLE thermostat lead to additional advantages.
For instance, we can contrast its behavior with that of Nos\'e-Hoover (NH) chains, 
based on equations which are quadratic in $p$ (see Appendix~\ref{app:nose}).
As a consequence of the nonlinearity, the efficiency
of an NH chains thermostat for a multidimensional oscillator depends 
on the orientation of the eigenmodes relative to the cartesian axes, an
artefact which is absent in our case.

Having set the background, we now turn to the description of the various applications
of Eqs.~(\ref{eq:mark-sde}).

\subsection{Efficient canonical sampling}\label{sub:canonical}
We first discuss the design of a GLE which can optimally sample phase space.
In this case, the target stationary distribution is the canonical ensemble,
so the equations of motion need to satisfy the detailed-balance condition. 
Still, there is a great deal of freedom available in the choice of the autocorrelation kernel
or, equivalently, in the choice of $\mathbf{A}_p$ and $\mathbf{B}_p$ matrices.
These free parameters can be used to optimize the sampling efficiency.
To this end, we must first  define an appropriate merit function.
Standard choices are the autocorrelation times of the 
potential and total energy ($V$ and $H$ respectively):
\begin{equation}
\begin{split}
 \tau_V=&\frac{1}{\left<V^2\right>}\int_0^\infty \left<(V(t)-\left<V\right>)(V(0)-\left<V\right>)\right> \mathrm{d}t \\
 \tau_H=&\frac{1}{\left<H^2\right>}\int_0^\infty \left<(H(t)-\left<H\right>)(H(0)-\left<H\right>)\right> \mathrm{d}t.  \\
\end{split} \label{eq:taus}
\end{equation}
In the harmonic case, these can be readily computed in terms of correlation times
of $q^2$ and $p^2$ (see Appendix~\ref{app:corr-time}), and will depend on $\mathbf{A}_p$ and the
oscillator's frequency $\omega$. For example, one easily finds that in the white-noise limit,
with no additional degrees of freedom as in Eq.~(\ref{eq:langevin}),
\begin{equation}
 \tau_H\left(\omega\right)=\frac{1}{a_{pp}}+\frac{a_{pp}}{4\omega^2}, \quad
 \tau_V\left(\omega\right)=\frac{1}{2a_{pp}}+\frac{a_{pp}}{2\omega^2}, \quad
. \label{eq:tauh-wn}
\end{equation}
Both response times are constant in the high-frequency limit, and 
increase quadratically in the low-frequency extreme of the spectrum. 
For a given frequency one can choose $a_{pp}$ so as to minimize the correlation time - thus
enhancing sampling.
It should be noted that Eqs.~(\ref{eq:tauh-wn}) contain a ``trivial'' dependence on $\omega$,
as one expects that sampling a normal mode would require at least a time of the 
order of its vibrational period. 
One can thus define a renormalized $\kappa(\omega)=\left[\tau(\omega) \omega\right]^{-1}$
as a measure of the efficiency of the coupling. In the white-noise case, $\kappa=1$
for the optimally-coupled frequency ($\omega_H=a_{pp}/2$ and $\omega_V=a_{pp}$, respectively), 
and decreases linearly for lower and higher values of $\omega$.

While this result in itself provides a guide to choose a good value of the friction coefficient
in conventional (white-noise) Langevin dynamics, we can
enhance the value of $\kappa(\omega)$ over a broader frequency range, by using
a colored-noise SDE. If we want to obtain canonical sampling, 
the FDT has to hold, so that $\mathbf{C}_p=k_B T$.  We therefore consider
the entries of  $\mathbf{A}_p$ as the only independent parameters, since 
$\mathbf{B}_p$ is then determined by Eq.~(\ref{eq:free-cov}).

In practice, we set up a fitting procedure, in which we choose a set of frequencies $\omega_i$, 
distributed over a broad range $\left(\omega_{min},\omega_{max}\right)$.
For an initial guess for the thermostat matrix $\mathbf{A}_p$ 
we compute $\kappa(\omega)$ for each of these frequencies.
We then vary $\mathbf{A}_p$, so as to optimize
$\min_i \kappa(\omega_i)$, and aim at a sampling efficiency on the range 
$\left(\omega_{min},\omega_{max}\right)$ which is as high and frequency-independent 
as possible. 
We will discuss this fitting procedure in more detail in Section~\ref{sec:fitting}.

\begin{figure}
 \includegraphics{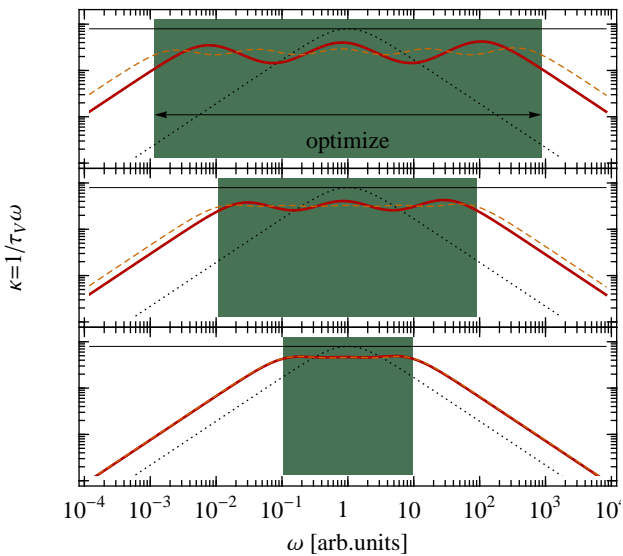}
\caption{\label{fig:gle-opt} Sampling efficiency as estimated from Eq.~(\ref{eq:taus}) 
for an harmonic oscillator, plotted as a function of the frequency $\omega$. The $\kappa(\omega)$
curve for a white-noise Langevin thermostat optimized for $\omega=1$ [black, dotted lines, Eq.(\ref{eq:tauh-wn})]
is contrasted with those for a set of optimized GLE thermostats. 
The panels, from bottom to top, contain the results fitted respectively over a frequency range spanning 
two, four and six orders of magnitudes around $\omega=1$. Dark, continuous lines correspond to 
matrices with $n=4$, and dashed, lighter lines to $n=2$.
The GLE curves correspond to the sets of parameters \protect\url{kv_4-2}, \protect\url{kv_2-2},
\protect\url{kv_4-4}, \protect\url{kv_2-2}, \protect\url{kv_4-6}, \protect\url{kv_2-6}, 
which can be downloaded from an on-line repository\cite{PARS}.
}
\end{figure}

In Figure~\ref{fig:gle-opt} we compare the optimized $\kappa(\omega)$ for different frequency ranges and 
number of additional degrees of freedom. We find empirically that $\kappa(\omega)=1$ is the 
best result which can be attained, and that nearly-optimal efficiency can be reached over a very broad
range of frequencies. This constant efficiency decreases slightly as 
the fitted range is extended, regardless of the number $n$ of $s_i$ employed.
For a given frequency range, however, increasing $n$ has the effect of making
the response flatter.

Clearly this scheme will work optimally in harmonic or quasi-harmonic systems,
and anharmonicity will introduce deviations from the predicted behavior.
In the extreme case of diffusive systems such as liquids, one has to ask the question of 
how much diffusion will be affected by the thermostat, especially since in 
an overdamped LE equation the diffusive modes are considerably slowed down
(see e.g.~Ref.~\cite{buss-parr08cpc}).
To estimate the impact of the thermostat on the diffusion, we define the
free-particle diffusion coefficient $D^*$ as that calculated switching off the
physical forces. Its value when a GLE thermostat is used is
\begin{equation}
\begin{split}
 \frac{m D^*}{k_B T}=&\frac{1}{\left<p^2\right>} \int_0^\infty \left<p(t)p(0)\right>\mathrm{d}t=\\
 =&\left[\mathbf{A}_p^{-1}\right]_{pp}=\left(a_{pp} -\mathbf{a}_p^T\mathbf{A}^{-1}\bar{\mathbf{a}}_p\right)^{-1}.
\end{split}
\label{eq:diff-coeff}
\end{equation}
where we have assumed the FDT to hold.
In practical cases, if an estimate of the unthermostated (intrinsic) diffusion coefficient $D$ is available,
one should choose the matrix $\mathbf{A}_p$ in such a way that $D^*\gg D$, so that the thermostat will not
behave as an additional bottleneck for diffusion.
Equation~\eqref{eq:diff-coeff} has the interesting consequence that $D^*$ can be enhanced either 
by reducing the overall strength of the noise, as in white-noise LE,
but also by carefully balancing the terms in the denominator of Eq.~(\ref{eq:diff-coeff}).

We have found empirically that for an $\mathbf{A}_p$ matrix fitted to harmonic modes 
over the frequency range $\left(\omega_{min},\omega_{max}\right)$, the diffusion 
coefficient computed by~(\ref{eq:diff-coeff}) is  $D^*\approx k_BT/\omega_{min}$. 
This latter expression gives a useful recipe for choosing the minimal frequency 
to be considered when fitting a GLE thermostat for a system whose diffusion coefficient 
can be roughly estimated.

\subsection{Frequency-dependent thermostatting}\label{sub:quantum}
\begin{figure}
 \includegraphics{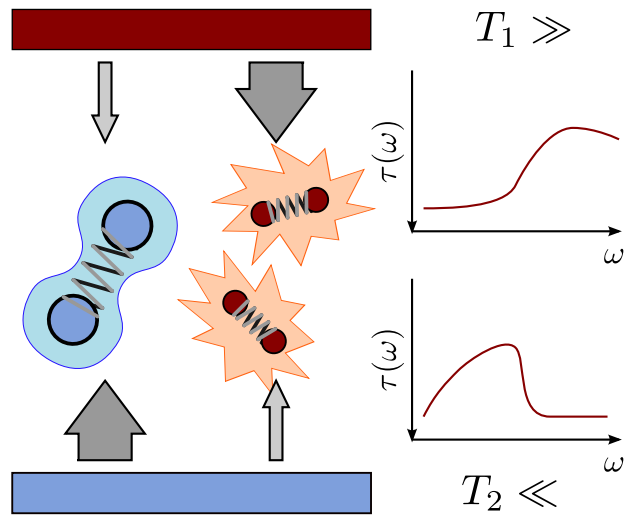}
\caption{\label{fig:twothermo} 
A cartoon representing a two-thermostat setup, which we take as the simplest 
example of a stochastic process violating the fluctuation-dissipation theorem.
If the relaxation time versus frequency curves for the two thermostats are different,
a steady-state will be reached in which normal modes corresponding to different frequencies 
will equilibrate at different effective temperatures.
}
\end{figure}

The ability to control the strength of the thermostat-system coupling as a function
of the frequency -- demonstrated above --  points quite naturally at 
more sophisticated applications.
For instance, one can apply two thermostats with 
distinct target temperatures and different efficiencies $\kappa(\omega)$ (see Figure~\ref{fig:twothermo}).
Obviously, such a simulation is not an equilibrium one, since energy is systematically
injected in some modes and removed from others,
but leads to a steady state that has useful properties.
Indeed, the normal modes will couple differently to the two thermostat, so that 
the effective temperature of each mode can be controlled as a function of $\omega$. 
This two-thermostats example is just an instance of a broader class of stochastic processes,
for whom the FDT is violated. In general, we can relax the assumption that 
$\mathbf{C}_p=k_B T$, and for a given drift matrix we can choose a 
$\mathbf{B}_p$ which is suitable to our purpose.

Returning to the harmonic oscillator case, one can solve exactly the dynamics
for a given choice of $\mathbf{A}_p$, $\mathbf{B}_p$ and frequency $\omega$.
The resulting dynamics is performed in the $n+2$-dimensional space defined by the
variables $(q,p,\mathbf{s})$, according to Eq.~(\ref{eq:mark-pq}).
For a compact notation, we used the full matrices $\mathbf{A}_{qp}$ and $\mathbf{B}_{qp}$.
The full $\mathbf{C}_{qp}(\omega)$, which defines the stationary distribution
in the steady state, can be computed solving an equation 
analogous to~(\ref{eq:free-cov}):
\begin{equation}
\mathbf{A}_{qp}\mathbf{C}_{qp}+\mathbf{C}_{qp}\mathbf{A}_{qp}^T=\mathbf{B}_{qp}\mathbf{B}_{qp}^T
\end{equation}
One can tune the free parameters ($\mathbf{A}_p$ and $\mathbf{B}_p$) so as to make the
$c_{qq}(\omega)$ and $c_{pp}(\omega)$ elements of the extended covariance matrix 
as close as possible to the desired target functions $\left<q^2\right>(\omega)$ 
and $\left<p^2\right>(\omega)$.

In a previous paper \cite{ceri+09prl2} we applied this method to obtain $\left<q^2\right>(\omega)$ 
and $\left<p^2\right>(\omega)$ in agreement with the values appropriate for 
a quantum harmonic oscillator, and obtained a good approximation to the
quantum-corrected structural properties in quasi-harmonic systems.
Many other applications can be envisaged, which take advantage of frequency-dependent
thermostatting. For instance, one could use this technique in  
accelerated sampling methods\cite{ross+02jcp,vand-roth02jpcb,mara-vand06cpl}, 
which work by artificially heating the low-frequency modes, whilst keeping the other modes
at the correct temperature.

\subsection{Implementation}\label{sub:implementation}
The implementation of a GLE thermostat in molecular-dynamics 
simulations is straightforward. Here, we consider the case of a velocity-Verlet 
integrator, which updates positions and 
momenta by a time step $\Delta t$, according to the scheme:
\begin{equation}
 \begin{split}
  p\leftarrow & p + V'(q) \Delta t /2\\
  q\leftarrow & q + p \Delta t\\
  p\leftarrow & p + V'(q) \Delta t /2.\\
 \end{split}
\label{eq:vv-integrator}
\end{equation}
Eqs.~(\ref{eq:vv-integrator}) can be obtained using Trotter
splitting in a Liouville operator formalism\cite{tuck+92jcp}.
In the same spirit one can introduce our GLE thermostat by performing two 
free-particle steps by $\Delta t/2$ on the $(p,\mathbf{s})$ variables\cite{buss-parr07pre}:
\begin{equation}
 \begin{split}
  \left(p,\mathbf{s}\right)&\leftarrow \mathcal{P}\left[\left(p,\mathbf{s}\right),\Delta t/2\right]\\
  p&\leftarrow p + V'(q) \Delta t /2\\
  q&\leftarrow q + p \Delta t\\
  p&\leftarrow p + V'(q) \Delta t /2.\\
\left(p,\mathbf{s}\right)&\leftarrow \mathcal{P}\left[\left(p,\mathbf{s}\right),\Delta t/2\right]\\
 \end{split}
\label{eq:gle-integrator}
\end{equation}

At variance with thermostats based on second-order equations of motion such as 
Nos\'e-Hoover, where a multiple time-step approach is required to obtain accurate trajectories\cite{tuck+96jcp,jang-voth97jcp},
this free-particle step can be performed without introducing additional sampling errors.
The exact finite-time propagator for $(p,\mathbf{s})$ reads:
\begin{equation}
\mathcal{P}\left[\left(p,\mathbf{s}\right),\Delta t\right]^T=
\mathbf{T}(\Delta t)\left(p,\mathbf{s}\right)^T
+\mathbf{S}(\Delta t)\boldsymbol{\xi}^T
\label{eq:fp-integrator}
\end{equation}
where $\boldsymbol{\xi}$ is a vector of $n+1$ uncorrelated Gaussian numbers, and the
matrices $\mathbf{T}$ and $\mathbf{S}$ can be computed once, at the beginning of the
simulation and for all degrees of freedom\cite{gard03book,fox+88pra}. 
The relations between $\mathbf{T}$, $\mathbf{S}$, $\mathbf{A}_p$, $\mathbf{C}_p$ 
and $\Delta t$ read
\begin{equation*}
\mathbf{T}=e^{-\Delta t \mathbf{A}_p}, 
\mathbf{S}\mathbf{S}^T=\mathbf{C}_p-e^{-\Delta t \mathbf{A}_p}\mathbf{C}_p e^{-\Delta t \mathbf{A}_p^T}.
\end{equation*}

It is worth pointing out that when FDT holds, the canonical distribution is invariant under the action
of~(\ref{eq:fp-integrator}), whatever the size of the time-step. A useful consequence of this property is that,
in the rare cases where applying~(\ref{eq:fp-integrator}) introduces a significant overhead over
the force calculation, the thermostat can be applied every $m$ steps of dynamics, using a stride
of $m\;\Delta t$.  This will change the trajectory, but does not affect the accuracy of sampling.

The velocity-Verlet algorithm~(\ref{eq:vv-integrator}) introduces finite-$\Delta t$ errors, whose 
effect needs to be monitored. In microcanonical simulations, this is routinely done by checking 
conservation of the total energy $H$. 
Following the work of Bussi~{\em et al.}\cite{buss+07jcp} we introduce a conserved quantity
$\tilde{H}$, which can be used to the same purpose:
\begin{equation}
 \tilde{H}=H-\sum_i \Delta K_i \label{eq:conserved}
\end{equation}
where $\Delta K_i$ is the change in kinetic energy due to the action of the thermostat at 
the $i$-th time-step, and the sum is extended over the past trajectory.
In cases where the FDT holds, such as that described in Section~\ref{sub:canonical}, the drift of the
effective energy quantitatively measures the violation of detailed balance induced
by the velocity-Verlet step, similarly to Refs.~\cite{buss+07jcp,buss-parr07pre}.
In the cases where the FDT does not hold, such as the frequency dependent
thermostating described in Section~\ref{sub:quantum}, the conservation of this quantity just measures
the accuracy of the integration, similarly to Refs.~\cite{brun+07jpcb,ensi+07jctc}.

\section{Fitting of colored-noise parameters}\label{sec:fitting}
A key feature of our approach resides in our ability to optimize the performance
of the thermostat based on analytical estimates, making the method effectively parameterless. 
Such optimization, however, is not trivial, even if computationally inexpensive. 
The relationship between $\mathbf{A}_p$, $\mathbf{B}_p$ and the correlation properties of 
the resulting trajectory is highly nonlinear. Furthermore, we found empirically that many local 
minima exist which greatly hinder the optimization process. 
With these difficulties in mind, we provide a downloadable library of fitted parameters\cite{PARS} 
which can be adapted to most of the foreseeble applications, according to the prescriptions 
given in Section~\ref{sub:scaling}.
Details about the fitting procedure are given in the following three subsections.

\subsection{Parameterization of GLE matrices}
A number of constraints must be enforced on the drift and diffusion matrices in order 
to guarantee that the resulting SDE is well-behaved. It is therefore important to find a 
representation of the matrices such that during fitting these conditions are automatically enforced,
and that the parameters space  is efficiently explored.
A first condition, required to yield a memory kernel with exponential decay, is that all the eigenvalues 
of $\mathbf{A}_p$ must have positive real part. A second requirement is that the kernel
$K(\omega)$ is positive for all real $\omega$. This ensures that the stochastic process 
will be consistent with the second law of thermodynamics\cite{ford+88pra}.

Finding the general conditions for $\mathbf{A}_p$ to satisfy this second constraint is not simple.
However, we can state that a sufficient condition for $K(\omega)>0$ is that 
$\mathbf{A}_p+\mathbf{A}_p^T$ is positive definite.
For simplicity we shall assume such a positivity condition to hold, since we found 
empirically that this modest loss of generality does not significantly affect the
accuracy or the flexibility of the fit.
Moreover, in the case of canonical sampling, $\mathbf{A}_p+\mathbf{A}_p^T>0$
is also required in order to obtain a real diffusion matrix, since 
$\mathbf{B}_p\mathbf{B}_p^T=k_BT\left(\mathbf{A}_p+\mathbf{A}_p^T\right)$ 
according to Eq.~(\ref{eq:free-cov}).

One would like to find a convenient parameterization, which enforces 
automatically these constraints. This is best done by writing 
$\mathbf{A}_p=\mathbf{A}_p^{(S)}+\mathbf{A}_p^{(A)}$, the sum of a symmetric and antisymmetric
part. Since any orthogonal transform of the $\mathbf{s}$ degrees of freedom would not
change the dynamics (see Appendix~\ref{app:memory}), one can assume without loss of 
generality that the $\mathbf{A}^{(S)}$ block in $\mathbf{A}_p^{(S)}$ is diagonal 
(see Eq.~(\ref{eq:notation}) for the naming convention).
Since in the general case the antisymmetric $\mathbf{A}_p^{(A)}$ does 
not commute with $\mathbf{A}_p^{(S)}$, we will assume it to be full,
while $\mathbf{A}_p^{(S)}$ can be written in the form
\begin{equation}
 \mathbf{A}_p^{(S)}=
\!\left(\!\begin{array}{ccccc}
a      & a_1 & a_2 & \cdots & a_n \\
a_1     & \alpha_1 & 0   & \cdots & 0   \\
a_2     & 0   & \alpha_2 & \ddots & 0   \\
\vdots & \vdots & \ddots  & \ddots & \vdots \\  
a_n    & 0 & 0 &\cdots & \alpha_n 
\end{array}\!\right)\!.
\end{equation}
In order to enforce the positive-definiteness, one use an analytical Cholesky decomposition 
$\mathbf{A}_p^{(S)}=\mathbf{Q}_p\mathbf{Q}_p^T$, with $\mathbf{Q}_p$ 
\begin{equation}
 \mathbf{Q}_p=
\!\left(\!\begin{array}{ccccc}
q      & q_1 & q_2 & \cdots & q_n \\
0     & d_1 & 0   & \cdots & 0   \\
0     & 0   & d_2 & \ddots & 0   \\
\vdots & \vdots & \ddots  & \ddots & \vdots \\  
0    & 0 & 0 &\cdots &d_n 
\end{array}\!\right)\!
\label{eq:fit-q}
\end{equation}
and $\alpha_i=d_i^2$, $a_i=d_i q_i$, and $a=q^2+\sum_i q_i^2$.
Such a parameterization guarantees that $\mathbf{A}_p$ will generate a dynamics with a stationary
probability distribution, and requires $2n+1$ parameters for the symmetric part 
(the elements of $\mathbf{Q}_p$, Eq.~(\ref{eq:fit-q})), and $n(n+1)/2$ for the antisymmetric 
part $\mathbf{A}_p^{(A)}$. If we want the equilibrium distribution to be the canonical,
one we must enforce the FDT, and $\mathbf{B}_p\mathbf{B}_p^T$ is uniquely determined. 

If we aim at a generalized formulation, which allows for frequency-dependent thermalization,
there are no constraints on the choice of $\mathbf{B}_p$ other than
the fact that both $\mathbf{B}_p\mathbf{B}_p^T$ and the covariance $\mathbf{C}_p$ must be positive-definite.
Clearly, a real, lower-triangular $\mathbf{B}_p$ is the most general parameterization of
a positive-definite $\mathbf{B}_p\mathbf{B}_p^T$, and amounts at introducing $(n+1)(n+2)/2$ 
extra parameters. Together with the assumption that $\mathbf{A}_p^{(S)}>0$, 
the condition $\mathbf{B}_p\mathbf{B}_p^T>0$ is sufficient to ensure that the unique symmetric $\mathbf{C}_p$ which 
satisfies~(\ref{eq:free-cov}) is also positive-definite.

\subsection{Fitting for canonical sampling}
Armed with such a robust and fairly general parameterization, 
one only needs to define a merit function to be optimized.
Again, we first consider the simpler case of canonical sampling. 
Here, we want to obtain a flat response over a wide, physically-relevant 
frequency range $\left(\omega_{min},\omega_{max}\right)$.
We have chosen the form
\begin{equation}
\chi_1=\left[\sum_i \left|\log \kappa(\omega_i)\right|^m\right]^{1/m},
\label{eq:chi-sampling}
\end{equation}
where $\omega_i$s are equally spaced on a logarithmic scale over the fitted range.
If a large value of $m$ is chosen, the $\omega_i$ which yields the 
lowest efficiency is weighted more, and a flat response curve is obtained.
We found empirically that values of $m$ larger than $~10$ lead to a 
proliferation of  local minima, and hinder efficient optimization.
To resolve this, one can use the optimal parameters for $m=2$ as input for further
refinement at larger $m$, until convergence is achieved.

\begin{figure}
 \includegraphics{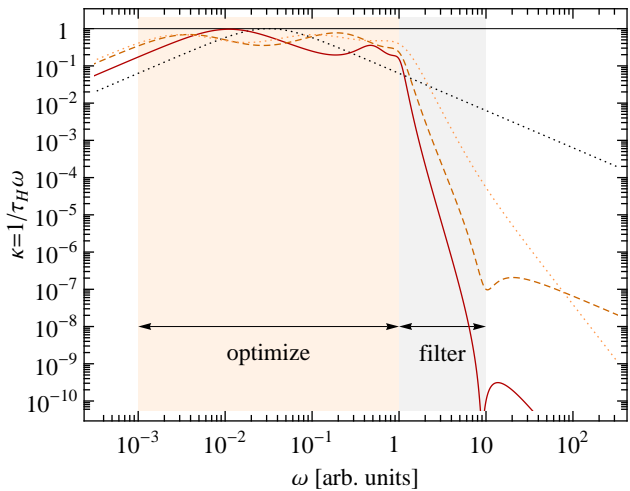}
\caption{\label{fig:gle-cpmd} Thermostatting efficiency, as estimated from Eq.~(\ref{eq:taus}),
for a colored-noise thermostat optimized for Car-Parrinello dynamics. Sampling efficiency
is optimized for $\omega\in (10^{-3},1)$, and an abrupt drop in efficiency is enforced
for $\omega\in (1,10)$, using the penalty function~(\ref{eq:chi-carparrinello}) 
in the fitting. The continuous (dark red) curve corresponds to $k=9$, the dashed (orange)
curve to $k=6$ and the dotted (light orange) curve to $k=3$. The $\kappa(\omega)$ curve 
for a white-noise thermostat centered on the optimized range is also reported for
reference (dotted black curve).
The three curves correspond to the parameters set \protect\url{cp-9_4-3}, \protect\url{cp-6_4-3} and 
\protect\url{cp-3_4-3}\cite{PARS}.
}
\end{figure}

This procedure can be modified so as to provide an efficient thermostat
which can be used in Car-Parrinello-like dynamics. In this case, the GLE
has to act as a low-pass filter in which only the low ionic frequencies
are affected, and fast electronic modes are not perturbed. To obtain this 
effect, we compute~(\ref{eq:chi-sampling}) only for the $\omega_i$'s
which are smaller than a cutoff frequency $\omega_{CP}$, and we introduce
an additional term
\begin{equation}
\chi_2=\left[\sum_{\omega_i>\omega_{CP}}
\left|\max\left[\log \kappa(\omega_i)- k (\omega_{CP}-\omega_i),0\right]\right|^m 
\right]^{1/m}.
\label{eq:chi-carparrinello}
\end{equation}
$\chi_2$ enforces a steep decrease of $\kappa(\omega)$ above $\omega_{CP}$, 
with a slope $k$ on a logarithmic scale. Values of $k$ as large as $9$ can 
be used, which guarantee an abrupt drop in thermalization efficiency 
for the fast modes (see Figure~\ref{fig:gle-cpmd}).

\subsection{Non-thermal noise and quantum thermostat}
We now discuss the case in which the thermostat is permitted
to violate FDT, in order to achieve frequency-dependent equilibration.
For these applications, one must also fit the fluctuations $c_{pp}(\omega)$ 
and $c_{qq}(\omega)$ to some target function $\tilde{c}_{pp}$ and $\tilde{c}_{qq}$.
We shall not treat the general case, but rather investigate the example
of the quantum thermostat (Ref.~\cite{ceri+09prl2}).
The procedure followed provides a clear guide for future extensions to different applications.

In order to reproduce quantum ions effects, one must selectively heat 
high-frequency phonons, for which zero-point energy effects are important, 
without affecting the low-frequency modes which behave classically.
The required frequency dependence of the variance for this case is
that of a quantum oscillator, i.e.
 $\tilde{c}_{pp}(\omega)=\omega^2 \tilde{c}_{qq}(\omega)=
\frac{\hbar\omega}{2} \coth\frac{\hbar\omega}{2 k_B T}$
The $\omega\rightarrow 0$, classical limit can be proved to correspond 
to two conditions on the elements of the free-particle covariance matrix
$\mathbf{C}_p$; namely, $c_{pp}=k_B T$ and $\mathbf{a}_p^T\mathbf{A}^{-1}\mathbf{c}_p=0$.
One could enforce such constraints exactly, by considering the entries of
 $\mathbf{C}_p$ as independent fitting parameters, and obtaining the 
diffusion matrix from Eq.~(\ref{eq:free-cov}).  We found however that this
choice makes it difficult to obtain a positive-definite $\mathbf{B}_p\mathbf{B}_p^T$, 
and that the fitting becomes more complex and inefficient.

As an alternative, we decided to enforce the low-frequency limit 
with an appropriate penalty function,
\begin{equation}
\chi_3=(c_{pp}/k_B T-1)^2+\left(\mathbf{a}_p^T\mathbf{A}^{-1}\mathbf{c}_p/k_B T\right)^2,
\label{eq:chi-climit}
\end{equation}
to be optimized together with the sampling efficiency~(\ref{eq:chi-sampling})
and a term which measures how well the finite-frequency fluctuations were fitted:
\begin{equation}
\chi_4=\left[\sum_i \left|\log \frac{c_{qq}(\omega_i)}{\tilde{c}_{qq}(\omega_i)} \right|^m+
\left|\frac{\log c_{pp}(\omega_i)}{\tilde{c}_{pp}(\omega_i)} \right|^m \right]^{1/m}
\label{eq:chi-fdt}
\end{equation}
Since the low-frequency limit is already enforced by~(\ref{eq:chi-climit}), we
compute~(\ref{eq:chi-fdt}) on a set of points equally spaced between the 
maximum frequency $\omega_{max}$ and one half of the onset frequency for
quantum effects $\omega_q=k_BT/\hbar$.

\subsection{Transferability of fitted parameters}\label{sub:scaling}

The scheme described in the previous Sections allowed us to obtain matrices 
suitable for all the  applications discussed in previous works. Furthermore, it provides 
a starting point for obtaining matrices which one might deem useful for novel applications.
However, the reader is advised that the fitting is still far from being
a black-box procedure. It is thus necessary to experiment with a combination of different initial
parameters and minimization schemes. We found the downhill simplex methods\cite{neld-mead65cj} to 
be particularly effective, but resorted to simulated annealing when the optimization 
got stuck in a local minimum.
There is a great deal of arbitrariness in the choice of the terms
(\ref{eq:chi-sampling}-\ref{eq:chi-fdt}), and in their weighted combination $\chi=\sum w_i \chi_i$. 
To make the procedure even more delicate, we observe that in high-$n$ cases the parameters
tend to collapse into ``degenerate'' minima, where the full dimensionality of the 
search space is not exploited. This phenomenon can be successfully 
circumvented by enforcing an even spacing of the eigenvalues of $\mathbf{A}$ over the 
frequency range of interest, and slowly releasing this restraint during the later stages of 
optimization.

However, the problems mentioned above have no major practical consequences, as the computation of analytical
estimates is inexpensive and one can afford a great deal of trial-and-error during the optimization. 
Moreover, fitted parameters can be reused, since the optimized parameters can be easily transferred
to similar problems because of the scaling properties of the dynamics~(\ref{eq:mark-pq}).

In fact, one can see that if the drift and covariance matrices 
$(\mathbf{A}_p,\mathbf{C}_p)$ lead to the efficiency curves $\kappa(\omega)$ and 
fluctuations $c_{pp}(\omega)$, the scaled matrices
$(\alpha\, \mathbf{A}_p,\beta\, \mathbf{C}_p)$ will yield $\kappa(\alpha^{-1}\omega)$,
and the fluctuations $\beta c_{pp}(\alpha^{-1}\omega)$.
This means that if $\mathbf{A}_p$ is optimized for sampling over the range $\left(\omega_{min},\omega_{max}\right)$,
$\alpha\, \mathbf{A}_p$ will be optimal over $\left(\alpha\,\omega_{min},\alpha\,\omega_{max}\right)$.
We also remark that if $(\mathbf{A}_p,\mathbf{C}_p)$ are fitted to the quantum harmonic oscillator
fluctuations at temperature $T$, $(\alpha\, \mathbf{A}_p,\alpha\, \mathbf{C}_p)$ will
be suitable for temperature $\alpha\,T$. Care must be taken in this case 
to ensure that the scaled frequency range still encompasses the whole vibrational
spectrum of the system being studied. 

\section{Understanding the quantum thermostat}

As discussed in Ref.\cite{ceri+09prl2}, one must pay a great deal of attention
when using a ``quantum thermostat'', because energy is transferred between
modes of different frequency, as a consequence of the anharmonic coupling.
This is reminiscent of zero-point energy (ZPE) leakage which plagues 
semiclassical approaches to the computation of nuclear quantum effects\cite{alim+92jcp,habe-mano09}.
In the cases we explored so far, empirical evidence suggests that quasi-harmonic solids,
can be treated with good accuracy down to temperatures as low as $10$\% of the Debye temperature 
$\Theta_D$. Clearly, the ultimate test to assess of the accuracy of the method is 
a comparison with path-integral calculations, to be performed on a similar but computationally cheaper
model, such as a smaller-size box or a simpler force field.

One would like however to obtain some qualitative measure of the quality of the fit,
and to gauge the transferability of a given set of parameters. To this end, we 
first state a couple of empirical rules, and then validate them on 
two fairly different real systems.
A first observation is that it is useless to push the fitting of the 
fluctuations $c_{pp}(\omega)$ and $c_{qq}(\omega)$ to very high accuracy, if 
this comes at the expense of the coupling efficiency. In fact, we would be trading 
a small, controlled fitting error with a possibly larger, uncontrollable and 
system-dependent error stemming from anharmonicity.
 Secondly, we observed that in order to contrast more effectively the flow of energy between different
phonons, one should try to reduce the correlation time of the kinetic energy $\tau_K$, rather than
focus solely on the terms~(\ref{eq:taus}), which are better suited to measure sampling efficiency. 
In fact, a low $\tau_K(\omega)$ corresponds to a slightly overdamped regime, 
where sampling efficiency is sub-optimal, but ZPE is enforced more tightly.

\begin{figure}
 \includegraphics{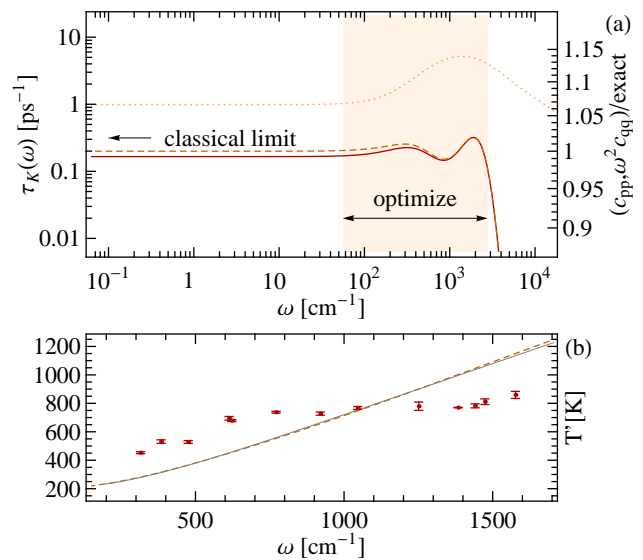}
\caption{\label{fig:qd-bad} 
(a): $\omega$-dependence of the kinetic energy correlation time $\tau_k(\omega)$ (light, dotted line) and
of the ratio of the fitted fluctuations $c_{pp}(\omega)$ (dashed line) and of $\omega^2 c_{qq}(\omega)$ (full line)
with the exact, quantum-mechanical target function.
(b): normal-mode-projected kinetic temperature for a few, selected phonons. The dashed line is the 
value expected from the fitting $c_{pp}(\omega)$, while the full line is the exact, quantum-mechanical 
expectation value for a harmonic oscillator.
Calculations have been performed with the parameters \protect\url{qt-20_6_BAD}\cite{PARS}.
}
\end{figure}

\begin{figure} 
 \includegraphics{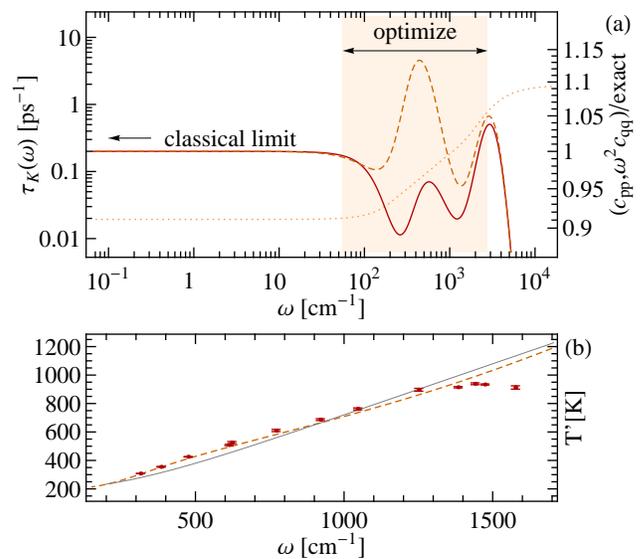}
\caption{\label{fig:qd-good} 
(a): $\omega$-dependence of the kinetic energy correlation time $\tau_K(\omega)$ (light, dotted line) and
of the ratio of the fitted fluctuations $c_{pp}(\omega)$ (dashed line) and of $\omega^2 c_{qq}(\omega)$ (full line)
with the exact, quantum-mechanical target function.
(b): normal-mode-projected kinetic temperature for a few, selected phonons. The dashed line is the 
value expected from the fitting $c_{pp}(\omega)$, while the full line is the exact, quantum-mechanical 
expectation value for a harmonic oscillator.
Calculations have been performed with the parameters \protect\url{qt-20_6}\cite{PARS}.
}
\end{figure}

To demonstrate these concepts in a real system, we performed some calculations with
a Tersoff model of diamond at a temperature $T=200$~K. At this low temperature,
slightly below $0.1 \Theta_D$, quantum effects are very strong, and we therefore expect to 
have problems maintaining the large difference in temperature between the stiff and 
soft phonons. Using a very harmonic system such as diamond is particularly useful,
since one can monitor directly the efficiency of the thermostat by 
projecting the atomic velocities on a selection of normal modes. Hence, a projected kinetic temperature
$T'(\omega)$ can be computed, and its value checked against the predictions in the harmonic limit, 
in the same spirit as in Ref.\cite{ceri+09prl2}.
In Figure~\ref{fig:qd-bad} we report the results with a matrix fitted taking into account
only the terms~(\ref{eq:chi-climit}) and~(\ref{eq:chi-fdt}). Even in a harmonic system 
such as diamond there are major errors due to ZPE leakage from the high-frequency
to the low-frequency modes, which the thermostat compensates only partially.
These poor results should be compared with those of Figure~\ref{fig:qd-good}. 
Here, we have also introduced in the fit a term analogous to~(\ref{eq:chi-sampling}) 
to reduce the value of $\tau_K(\omega)$. The projected kinetic temperature now agrees 
almost perfectly with the analytical predictions $c_{pp}(\omega)$ for most of the modes.
The only ones displaying significant deviations are the faster ones, 
for whom the value of $\tau_K(\omega)$ is slightly larger.
The $c_{pp}(\omega)$ curve deviates by nearly $10$\% from the exact, quantum-mechanical
expectation value. However, thanks to the more efficient coupling, the errors due to 
anharmonicities are better compensated, and in actuality, the overall error is much smaller than 
for the parameters presented in Figure~\ref{fig:qd-bad}.

\begin{figure}
 \includegraphics{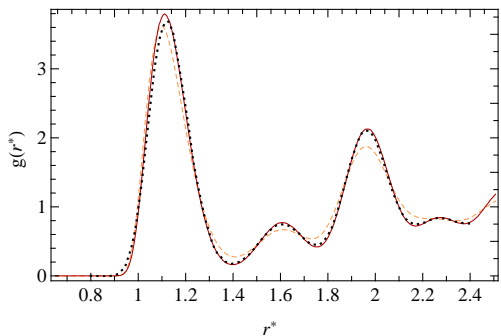}
\caption{\label{fig:qneon} 
Radial distribution function as computed from fully-converged path-integral calculations\cite{sing88molp}
(black, dotted line), and from a quantum-thermostat MD trajectory for a Lennard-Jones model 
of solid neon at $T=20$~K. Distances are in reduced units. Full line corresponds to 
the parameters set \protect\url{qt-20_6} (cfr. Figure~\ref{fig:qd-good}), and 
lighter, dashed line to the set \protect\url{qt-20_6_BAD} (cfr. Figure~\ref{fig:qd-bad}).
}
\end{figure}

To test whether these prescriptions work for less harmonic problems, we now turn 
to a completely different system; namely, the structural properties of solid neon at $20$~K.
At variance with diamond, quantum-ions effects are less pronounced, but the system
is close to its melting temperature, and it is significantly anharmonic.
As shown in Figure~\ref{fig:qneon}, the agreement between our results and those of
accurate path-integral calculations\cite{sing88molp} is almost perfect
if the parameters of Figure~\ref{fig:qd-good} are used. As expected, large errors are present if
\protect\url{qt-20_6_BAD} is used.
Further improvements on the fitting strategy, and the application to strongly 
anharmonic systems is currently being investigated, and will be the subject
of further work.

\section{Conclusions}
In this paper we have discussed in detail the use of colored-noise dynamics,
based on Ornstein-Uhlenbeck processes, as a tool for performing molecular dynamics.
Applications range from enhanced sampling, which we demonstrate in the harmonic limit
and which will be applied to real systems in forthcoming publications, to thermostats
for adiabatically separated problems and frequency-dependent thermalization.

Our idea exploits the linear nature of the OU stochastic differential equations,
which allows one to use the one-dimensional harmonic oscillator as a simple but
physically-motivated reference model. On the basis of the analytical prediction obtained
in that case, we describe a recipe for fitting the thermostat parameters so as to 
obtain the desired response properties in real systems.
The procedure is not simple, and we are considering different 
approaches to make it more robust and effective. Fortunately
however, fitted matrices can be easily transferred from one system to another.
With this in mind we have provided an extensive library of optimized parameters\cite{PARS},
which makes fitting unnecessary for most applications.

We also comment on practical 
issues concerning the implementation of the generalized-Langevin thermostat
in a molecular-dynamics program and its use in applications.
In particular, we discuss in detail how one can use colored noise
to model nuclear quantum effects\cite{ceri+09prl2}.
We provide some empirical rules to guide the fitting in this difficult case,
and we demonstrate that a normal-mode analysis in a quasi-harmonic system
is a valuable tool for assessing the quality of a set of parameters.
We believe that further investigation will find many other applications for
colored-noise in molecular-dynamics, and in computer simulations of 
molecular systems in general. As an example, we are currently investigating using
a zero-temperature, optimal-sampling GLE thermostat in order to perform structural
optimization. On similar lines, and taking inspiration from ``quantum annealing''\cite{lee-bern00jpca,sant-tosa06jpa}, 
one can envisage using frequency-dependent thermalization to improve the performance of simulated annealing.

\section{Acknowledgements}
The authors would like to thank David Manolopoulos for important suggestions
and fruitful discussion, Marcella Iannuzzi for having implemented the GLE 
thermostat in CP2K\cite{CP2K}, Alessandro Curioni for implementation in CPMD\cite{CPMD},
Grigorios Pavliotis and Michela Ottobre for discussion and references on 
stochastic processes. We also acknowledge Stefano Angioletti-Uberti, Paolo Elvati, 
Hagai Eshet, Kuntal Hazra,  Rustam Khaliullin and  Tom Markland for discussion and
for having preliminarily tested the thermostat in real applications, providing 
valuable feedback. We are especially in debt with Gareth Tribello, 
who contibuted to testing and greatly helped us improving the manuscript.

\appendix
\section{Memory kernels for the non-Markovian formulation}
\label{app:memory}
The connection between the Markovian~(\ref{eq:mark-sde}) and non-Markovian~(\ref{eq:nonmark-sde})
formulations of the colored-noise Langevin equation can be understood using techniques similar to those 
adopted in Mori-Zwanzig theory\cite{zwan+01book,lucz05chaos}.
Let us first consider a very general, multidimensional OU process,
where we single out some degrees of freedom ($\mathbf{y}$) that we wish to integrate out,
leaving only the variables marked as $\mathbf{x}$. 
\begin{equation}
 \label{eq:dred-mark}
\!\left(\!
\begin{array}{c}
 \dot{\mathbf{x}} \\\hline
 \dot{\mathbf{y}} \\
\end{array}
\!\right)\! 
=
-\!\left(\!
\begin{array}{c|c}
 \mathbf{A}_{xx} & \mathbf{A}_{xy} \\\hline 
 \mathbf{A}_{yx} & \mathbf{A}_{yy} \\
\end{array}
\!\right)\!
\!\left(\!
\begin{array}{c}
 \mathbf{x} \\\hline
 \mathbf{y} \\
\end{array}
\!\right)\!
+
\!\left(\!
\begin{array}{c}
\quad \mathbf{B}_{x\xi} \quad\quad \\\hline
\quad \mathbf{B}_{y\xi} \quad\quad \\
\end{array}
\!\right)\!
\!\left(\!
\begin{array}{c}
\multirow{2}{*}{$\boldsymbol{\xi}$}\\\\
\end{array}
\!\right)\!
\end{equation}

Assuming that the dynamics has finite memory, one can 
safely take $\mathbf{y}(-\infty)=0$, and the ansatz
\begin{equation}
 \mathbf{y}(t)=\int_{-\infty}^t\!\!\! e^{-(t-t')\mathbf{A}_{yy}} \left[-\mathbf{A}_{yx}\mathbf{x}(t') 
+\mathbf{B}_{y\xi} \boldsymbol{\xi}(t')\right] \mathrm{d}t'.
\label{eq:dred-ansatz}
\end{equation}
Substituting into (\ref{eq:dred-mark}), one sees that $\mathbf{y}$ can be eliminated from
the dynamics of $\mathbf{x}$, and arrives at
\begin{equation}
\begin{split}
 \dot{\mathbf{x}}(t)=&-\int_{-\infty}^t \mathbf{K}(t-t') \mathbf{x}(t')\mathrm{d}t'+
\boldsymbol{\zeta}(t)\\
\mathbf{K}(t)=&2\mathbf{A}_{xx}\delta(t)-\mathbf{A}_{xy} e^{-t\mathbf{A}_{yy}}\mathbf{A}_{yx}\quad\left(t\ge0\right)\\
\boldsymbol{\zeta}(t)=&\mathbf{B}_{x\xi} \boldsymbol{\xi}(t) -\int_{-\infty}^t \mathbf{A}_{xy} e^{-(t-t')\mathbf{A}_{yy}} \mathbf{B}_{y\xi} \boldsymbol{\xi}(t').
\end{split}
\label{eq:dred-nonmark}
\end{equation}
One can see that~(\ref{eq:dred-nonmark}) are invariant under 
any orthogonal transformation of the $\mathbf{y}$ dynamical variables, meaning that
such a transformation leaves the dynamics of the $\mathbf{x}$'s unchanged.

The colored noise is better described in terms of its time-correlation function,
$\mathbf{H}(t)=\left<\boldsymbol{\zeta}(t)\boldsymbol{\zeta}(0)^T\right>$. Let us first 
introduce the symmetric matrix $\mathbf{D}=\mathbf{B}\mathbf{B}^T$, whose parts
we shall label using the same scheme used for $\mathbf{A}$ in Eq.~(\ref{eq:dred-mark}).
We shall also need $\mathbf{Z}_{yy}=\int_0^\infty e^{-\mathbf{A}_{yy}t}\mathbf{D}_{yy} e^{-\mathbf{A}_{yy}^T t}\mathrm{d}t$.
With these definitions in mind, one finds
\begin{equation}
\mathbf{H}(t)=\delta(t)\mathbf{D}_{xx}+
\mathbf{A}_{xy} e^{-t\mathbf{A}_{yy}}\left[\mathbf{Z}_{yy}\mathbf{A}_{xy}^T-\mathbf{D}_{yx}\right]  \quad \left(t\ge0\right).
\label{eq:dred-ht}
\end{equation}
Note that the value of $\mathbf{H}(t)$ for $t<0$ is determined by the constraint $\mathbf{H}(-t)=\mathbf{H}(t)^T$;
the value of $\mathbf{K}(t)$ instead, is irrelevant for negative times: 
we will assume $\mathbf{K}(-t)=\mathbf{K}(t)^T$ to hold, since this will simplify some algebra below.

Let's now switch to the case of the free-particle counterpart of Eqs.~(\ref{eq:mark-sde}),
which is relevant to the memory functions entering Eqs.~(\ref{eq:nonmark-sde}). Here, we
want to integrate away all the $\mathbf{s}$ degrees of freedom, retaining only the momentum $p$.
Hence, we can transform Eqs.~(\ref{eq:dred-nonmark}) and~(\ref{eq:dred-ht}) to the less 
cumbersome form
\begin{equation}
\begin{split} 
K(t)=&2a_{pp} \delta(t)-\mathbf{a}_p^T e^{-\left|t\right|\mathbf{A}}\bar{\mathbf{a}}_p\\
H(t)=& d_{pp} \delta(t)-\mathbf{a}_p^T e^{-\left|t\right|\mathbf{A}}\left[\mathbf{Z}\mathbf{a}_p-\mathbf{d}_p\right]
\end{split}
\label{eq:dred-mem-t}
\end{equation}
This compact notation hides certain relevant property of the memory kernels, which are more
apparent when the kernels are written in their Fourier representation.
If $\mathbf{D}_p=\mathbf{B}_p\mathbf{B}_p^T$ is transformed according to Eq.~(\ref{eq:free-cov}). 
$K(\omega)$ and $H(\omega)$ read
\begin{equation}
\begin{split}
K(\omega)=& 2a_{pp}-2  \mathbf{a}_p^T \frac{\mathbf{A}}{\mathbf{A}^2+\omega^2}\bar{\mathbf{a}}_p \\
H(\omega)=& K(\omega) \left(c_{pp}-  \mathbf{a}_p^T \frac{\mathbf{A}}{\mathbf{A}^2+\omega^2}\mathbf{c}_p\right) +\\
          &\!\!\!\!\!\!\!\!\!\!\!\!\!\!\!\!\!\!+2\omega^2 \left(\mathbf{a}_p^T \frac{1}{\mathbf{A}^2+\omega^2}\mathbf{c}_p \right)  
           \!\! \left(1+\mathbf{a}_p^T  \frac{1}{\mathbf{A}^2+\omega^2} \bar{\mathbf{a}}_p \right).
\end{split}
\label{eq:dred-mem-w}
\end{equation}
It is seen that the memory functions (hence the dynamical trajectory) are independent of the value
of $\mathbf{C}$, the covariance of the fictitious degrees of freedom.  Moreover,
a sufficient condition for the FDT to hold is readily found. By setting
$c_{pp}=k_BT$ and $\mathbf{c}_p=0$, one obtains $H(\omega)=k_B T K(\omega)$, which is 
precisely the FDT for a non-Markovian Langevin equation. Since the value of 
$\mathbf{C}$ is irrelevant we can take $\mathbf{C}_p=k_B T$, which simplifies 
the algebra and leads to numerically-stable trajectories.

\section{Covariance matrix and correlation times for the harmonic oscillator}
\label{app:corr-time}
Given $\mathbf{A}$ and $\mathbf{C}$ matrices (the drift 
term and the static covariance for a generic OU process), 
one can find the diffusion matrix $\mathbf{B}$ 
by an expression analogous to Eq.~(\ref{eq:free-cov}). 
The same relation can be used to obtain the elements of $\mathbf{C}$
given the drift and diffusion matrices, by solving the linear system.
However, the covariance matrix can be computed more efficiently by 
finding the eigendecomposition of $\mathbf{A}=\mathbf{O}\;\mathrm{diag}(\alpha_i)\;\mathbf{O}^{-1}$, 
and computing
 \begin{equation}
 C_{ij}=
\sum_{kl} \frac{O_{ik}
\left[\mathbf{O}^{-1}\mathbf{B}\mathbf{B}^T {\mathbf{O}^{-1}}^T\right]_{kl}O_{jl}}
{\alpha_k+\alpha_l}. \label{eq:cov}
\end{equation}

Now, let $\mathbf{x}$ be the vector describing the trajectory of the OU 
process. In order to compute $\tau_H$ or $\tau_V$ (Eq.~(\ref{eq:taus})) one needs
time correlation functions of the form $\left<x_i(t)x_j(t) x_k(0)x_l(0)\right>$. 
The corresponding, non-normalized integrals
\begin{equation}
 \tau_{ijkl}=\int_0^\infty\left[ \left<x_i(t)x_j(t) x_k(0)x_l(0)\right> -\left<x_i x_j\right>\left<x_k x_l\right>\right] \mathrm{d}t
\end{equation}
can be computed in terms of the tensorial quantity
\begin{equation}
X_{ijkl}= \sum_{mn}
\frac{O_{im}\left[\mathbf{O}^{-1}\mathbf{C}\right]_{ml}O_{jn}\left[\mathbf{O}^{-1}\mathbf{C}\right]_{nk}}{\alpha_m+\alpha_n}
\label{eq:xijkl}
\end{equation}
as $\tau_{ijkl}=\frac{1}{4}\left(X_{ijkl}+X_{ijlk}+X_{klij}+X_{lkij}\right)$.
For example -- if we consider the full OU process in the harmonic case -- one computes
\begin{equation}
 \tau_H=\frac{\omega^4\tau_{qqqq}+2\omega^2\tau_{qqpp}+\tau_{pppp}}{\omega^4 c_{qq}^2 +2 \omega^2 c_{qp}^2+c_{pp}^2},\quad 
\tau_V=\frac{\tau_{qqqq}}{c_{qq}^2}
\end{equation}
where we use an obvious notation for the indices in $\tau_{ijkl}$.

\section{A comparison with Nos\'e-Hoover Chains \label{app:nose}}
The most widespread techniques for canonical sampling in MD are probably
white-noise Langevin and Nos\'e-Hoover chains (NHC). White-noise Langevin
can be considered as a limit case of the thermostatting method we describe in this work,
but NHC is based on a redically different philosophy. It is therefore worth performing a 
brief comparison between the latter and the GLE thermostat.

In the ``massive'' version of the NH thermostat\cite{nose84jcp,hoov85pra}, each component of the physical momentum
is coupled to an additional degree of freedom with a fictitious mass $Q$, by means of a second-order 
equation of motion. The resulting dynamics ensures that the physically-relevant degrees of freedom will sample the
correct, constant-temperature ensemble, with the advantage of having deterministic equations of 
motion, and a well-defined conserved quantity.  However, in the harmonic case, trajectories are
poorly ergodic. This problem can be addressed by coupling the fictitious momentum to a second
bath variable with a similar equation of motion. By repeating this process further a ``Nos\'e-Hoover chain''
can be formed, which ensures that the dynamics is sufficiently chaotic to achieve efficient sampling\cite{mart+92jcp,tuck+93jcp}. 
The drawback of this approach is that the thermostat equations are second-order in momenta. 
It is therefore difficult to obtain analytical predictions for the properties of the dynamics,
and the integration of the additional degrees of freedom must be performed with a multiple time-step
approach, which makes the thermostat more expensive. 

To examine the performances of NHC and GLE, one could envisage comparing the sampling efficiency 
as defined by the correlation times~(\ref{eq:taus}).
Obtaining such estimates is not straightforward, not only because the the harmonic case cannot be treated
analytically, but also because in the multidimensional case the properties of the trajectory will not be invariant 
under an orthogonal transformation of coordinates, as discussed in Section~\ref{sec:gle}.
 The simplest model we can conceive for comparing NHC and GLE is therefore a two-dimensional
harmonic oscillator, with different vibrational frequencies on the two normal modes 
and adjustable relative orientations of the eigenvectors with respect to the thermostatted coordinates.

\begin{figure}
 \includegraphics{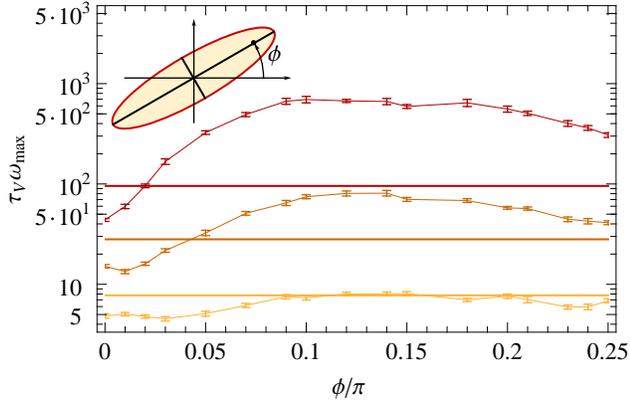}
\caption{\label{fig:rot-nose} Correlation time for the potential energy of a 2-D harmonic oscillator, as a function of
the angle between the eigenmodes and the cartesian axes. $\tau_V$ is computed for different values of 
the condition number $\omega_{max}/\omega_{min}$, from bottom to top 10, 31.6 and 100.
Thin lines serve as an aid for the eye, connecting the results obtained
in the three cases using a massive NH chains thermostat, with four additional degrees of freedom and
$Q=k_BT/\omega_{max}^2$. Error bars are also shown for individual data points.
Thick lines correspond to the (constant) result predicted for a GLE thermostat, 
using respectively the thermostat parameters \protect\url{kv_2-1}, centered on $0.32 \omega_{max}$,
 \protect\url{kv_4-2}, centered on $0.18 \omega_{max}$, and
\protect\url{kv_4-2} centered on $0.1 \omega_{max}$. The values obtained in
actual GLE simulations agree with the predictions within the statistical errorbar, 
and are not reported.
}
\end{figure}

The resulting $\tau_V$ is reported in Figure~\ref{fig:rot-nose}: 
in the highly anisotropic cases, the efficiency of the NH chains depends dramatically 
on the orientation of the axes, while for well-conditioned problems is almost constant.
The linear stochastic thermostat, on the other hand, has a predictable response, which 
is completely independent on orthogonal transforms of the coordinates.
In the one-dimensional case -- or when eigenvectors are perfectly aligned with axes --
NH chains are very efficient for all modes with frequency $\omega<\sqrt{\frac{k_BT}{Q}}$.
One should however consider that, in the absence of an exact propagator, 
choosing a small $Q$ implies that integration of the trajectory for the chains 
will become more expensive.

Obviously, such a simple toy model does not give quantitative information on the behavior 
in real-life cases, where modes of different frequencies coexist with anharmonicity 
and diffusive behavior. However, it demonstrates that the colored-noise Langevin 
thermostat performs almost as well as the axis-aligned NH chains. Furthermore, unlike the
NHC, there are no unpredictable failures for anisotropic potentials.

\end{document}